\documentclass[12,twocolumn]{article}
\usepackage{amsmath}
\usepackage{listings}
\usepackage{xcolor}
\usepackage{chngcntr}
\usepackage{url}

\begin{document}
\title{Quantum string comparison method}

\author{Vikram Menon \\
  Ayan Chattopadhyay$^{*}$}

\twocolumn[
  \begin{@twocolumnfalse}
    \maketitle
    \begin{abstract}
 We propose a quantum string comparison method whose main building blocks are a specially designed oracle construction followed by Grover's search algorithm. The purpose of the oracle is to compare all alphabets of the string in parallel. This requires a unique input state preparation, which when combined with some ancillas will result in a deterministic binary success and failure compare outcome.
    \end{abstract}
  \end{@twocolumnfalse}
]

\subsection*{Introduction}
String comparison is a basic ingredient of all the searching and sorting algorithms. Classical string comparison operate in $O(N)$, while the qunatum version can be built to exploit the superposition and parallelism inherent in the qunatum world. Here, we have devised one quantum string comparison method which constitutes oracle preparation with special input states and Grovers's search algorithm \cite{GRV}. The oracle preparation has a close resemblance with \cite{OR}, although the cited oracle operates in $O(N)$ using atleast $4$ qubits per bit comparison ($2$ input qubits to be compared and $2$ ancilla qubits) and universal single qubit and CNOT gates. In contrast, the proposed method would require $(N-1)*\log_{2}N$ ancilla qubits per $N$ bit comparison, i.e. $\frac{1}{2}$ ancilla per bit, which is almost 25\% ancillas per bit comparison. Here, we propose the oracle comparing an $N$ alphabet string to be split into $N$ sub oracles that can be operated in parallel, and thereby enhancing the overall performance. Apart from the similarity in oracle preparation, the overall method is different with respect to the input state preparation, number of ancilla qubits and the use of Grover's search. The method succeeds with 100\% probability yielding the result whether the strings are equal or not.

\section*{Oracle contruction}
Let us consider an input string $A$ of size $|A|=N$, over an alphabet set $\sum$. Here, each alphabets can be represented by $n$-qubits, where$n = \log_{2}N \Rightarrow 2^{n} = N$. A standard oracle would compare all or a subset of alphabets to perform the match, depending on the success probability desired ($O(N)$ to $O(\sqrt{N}$). We propose a modification by splitting this oracle into $N$ sub oracles, each capable of comparing a single alphabet ($i^{th}$ alphabet), say $A_{i} \in A$. The oracle sub unit can be defined as follows,
\begin{equation}
  O^{(A_{x},x)} = f_{A_{x}}(x) =
  \begin{cases}
    1, &         \text{if 'x' is a solution}\\
    0, &         \text{otherwise}.
  \end{cases}
\end{equation}
where, $x$ is a solution if the alphabet at postion $x$ is equal to the $x^{th}$ alphabet of the input string, i.e. $A_{x}$. The output of the these sub oracles are combined together using an $\mathrm{AND}$ gate to get the final output.
\begin{equation}
  \begin{split}
    O^{(A,x)} = &f_{A}(x) \\
    = &f_{A_{0}}(0) \; \mathrm{AND} \; f_{A_{1}}(1) \; \mathrm{AND}\\
    & \dotsb \; \mathrm{AND} \; f_{A_{N-1}}(N-1)
  \end{split}
\end{equation}
The sub oracles, $f_{A_{x}}(x)$, defined above can be run in parallel, making the amortized running time for oracle preparatiom $O(1)$, a considerable reduction of overall running time compared to a standard string comparison oracle.

\section*{Input state preparation}
Now, we will consider the input state preparation. Let's label the alphabets of the input string $A$ from $0$ to $N-1$, which will result in an $n$ qubit input state $|\psi\rangle$ of dimension $N$.
\begin{equation}
  \label{in_state}
  \begin{split}
    |\psi\rangle &= \frac{1}{\sqrt{N}} \sum_{x=0}^{N-1} |x\rangle\\
    &= \frac{1}{\sqrt{N}} (|0\rangle + |1\rangle + ... + |N-1\rangle)\\
    &= \frac{1}{\sqrt{N}} (|0_{1}0_{2}...0_{n}\rangle + |0_{1}0_{2}...1_{n}\rangle + ... + |1_{1}1_{2}...1_{n}\rangle)
  \end{split}
\end{equation}

Now, introduce additional $n$-qubits in the state $|1\rangle = |0_{1}0_{2}...1_{n}\rangle$ to the input,
\begin{equation}
  \begin{split}
    |\psi\rangle|\otimes|0_{1}0_{2}...1_{n}\rangle = \frac{1}{\sqrt{N}} (&|0\rangle|1\rangle + |1\rangle|1\rangle \\
    &+ ... + |N-1\rangle|1\rangle
  \end{split}
\end{equation}
Next add another $n$-qubits in state $|2\rangle$, then $|3\rangle$ and so on till $|N-1\rangle$, making it $((N-1)*n)$ ancillas. This will result in the following $nN$ qubit combined state.
\begin{equation}
  \label{comb_in_state}
  \begin{split}
    |\psi\rangle \otimes \Pi_{x=1}^{N-1} |x\rangle &= \\
    \frac{1}{\sqrt{N}} ( & \Pi_{x=0}^{N-1} |x\rangle + |1\rangle \otimes \Pi_{x=1}^{N-1} |x\rangle +\\
    & \dotsb + |N-1\rangle \otimes \Pi_{x=1}^{N-1} |x\rangle) = \\
    \frac{1}{\sqrt{N}} &(\sum_{x=0}^{N-1} |x\rangle) \otimes \Pi_{y=1}^{N-1} |y\rangle
  \end{split}
\end{equation}

The first basis state here, a special state, is a tensor product of all the $n$-qubit basis states $\Pi_{x=0}^{N-1}$ and forms the exact sequence of labels corresponding to the alphabets in the input string. All the other basis states have an incorrect first alphabet label.

\section*{Comparision method}
The oracle defined above, when operated on the input prepared in the previous section, will mark the first basis state if the strings to be compared are equal. This is because all the sub oracle functions will return $\forall i \in N: f_{A_{i}}(x_{i}) = 1$. For all the other basis states, the first sub oracle will return $0$.
\begin{equation}
  \begin{split}
    O^{(A,x)}[&\frac{1}{\sqrt{N}} (\Pi_{x=0}^{N-1}|x\rangle + |1\rangle\otimes\Pi_{x=1}^{N-1}|x\rangle \\
      &+ \dotsb + |N-1\rangle\otimes\Pi_{x=1}^{N-1}|x\rangle) =\\
      \frac{1}{\sqrt{N}} (&(-1)^{f_{A_{0}}(0) \; \mathrm{AND} \; f_{A_{1}}(1) \; \mathrm{AND} \; \dotsb \; \mathrm{AND} \; f_{A_{N-1}}(N-1)}\\
        &\Pi_{x=0}^{N-1}|x\rangle  +\\
        & (-1)^{f_{A_{0}}(1) \; \mathrm{AND} \; f_{A_{1}}(1) \; \mathrm{AND} \; \dotsb \; \mathrm{AND} \; f_{A_{N-1}}(N-1)}\\
        &|1\rangle\Pi_{x=1}^{N-1}|x\rangle \\
        & + \dotsb + \\
        & (-1)^{f_{A_{0}}(N-1) \; \mathrm{AND} \; f_{A_{1}}(1) \; \mathrm{AND} \dotsb \mathrm{AND} \; f_{A_{N-1}}(N-1)}\\
        &|N-1\rangle\Pi_{x=1}^{N-1}|x\rangle)
  \end{split}
\end{equation}

The Grover iterator, excluding the oracle, will operate only on the first $n$-qubits. Since the ancilla qubits are not entangled with the initial superposed state, there will not be any interference in the search operation.

The Grover's iterator will amplify the first basis state if the strings are equal, thereby transforming the input state given by equation \ref{comb_in_state} to $|0\rangle^{\otimes n} \otimes \Pi_{x=1}^{N-1} |x\rangle$. This can be confirmed by measuring the first $n$ input qubits, which will be in $|0\rangle^{\otimes n}$ state. In case of a mismatch, the first $n$-qubits will be unaltered and will remain in the equal superpostion state given by equation \ref{in_state};

\section*{Conclusion}
We have shown a quantum string comparison method built on the Grover's search algorithm. With the proposed oracle construction and a unique input state preparation, combining the input qubits with some  additional ancilla qubits, the comparison outcome was shown to be deterministic in $O(\sqrt{N})$. The outcome would be binary, zero if the strings match, non-zero otherwise.


\begin{thebibliography}{1}

\bibitem{GRV}
Grover, L., A fast quantum mechanical algorithm for database search. Proceedings of 28th ACM Symposium on Theory of Computing, 1996, pp. 212-219.  
\bibitem{OR}
Oliveira, D. and Ramos, R. (2007). Quantum bit string comparator: Circuits and applications. Quantum Computers and Computing, {\bf 7}.

\end{thebibliography}
\end{document}